\documentclass[journal]{IEEEtran}

\usepackage{url}
\usepackage{graphicx}
\usepackage{float}
\usepackage{amsmath}
\usepackage{amssymb}
\usepackage{hyperref}
\usepackage{xcolor}
\usepackage{booktabs}
\usepackage{multirow}
\usepackage{cite}
\usepackage{algorithm}
\usepackage{algorithmic}
\usepackage{subcaption}

\newcommand{\ket}[1]{|#1\rangle}
\newcommand{\bra}[1]{\langle#1|}
\newcommand{\commutator}[2]{[#1, #2]}

\begin{document}

\title{Constrained Portfolio Optimization via Quantum Approximate Optimization Algorithm (QAOA) with XY-Mixers and Trotterized Initialization: A Hybrid Approach for Direct Indexing}

\author{Javier~Mancilla$^{1*}$, Theodoros D. Bouloumis$^2$, Frederic Goguikian$^1$%
\thanks{1. SquareOne Capital, 1221 Brickell Ave. Suite 900, Miami Fl 33131., Miami, US

2. School of Physics, Faculty of Sciences, Aristotle University of Thessaloniki, GR-54124 Thessaloniki, Greece

Email Address: javier.m@squareonecap.com, tmpoulou@physics.auth.gr, fredi.g@squareonecap.com}
}

\maketitle

\begin{abstract}
Portfolio optimization under strict cardinality constraints is a combinatorial challenge that defies classical convex optimization techniques, particularly in the context of ``Direct Indexing'' and ESG-constrained mandates. In the Noisy Intermediate-Scale Quantum (NISQ) era, the Quantum Approximate Optimization Algorithm (QAOA) offers a promising hybrid approach. However, standard QAOA implementations utilizing transverse field mixers often fail to strictly enforce hard constraints, necessitating soft penalties that distort the energy landscape. This paper presents a comprehensive analysis of a constraint-preserving QAOA formulation against Simulated Annealing (SA) and Hierarchical Risk Parity (HRP). We implement a specific QAOA ansatz utilizing a Dicke state initialization $\ket{D_N^K}$ and an XY-mixer Hamiltonian that strictly preserves the Hamming weight of the solution, ensuring only valid portfolios of size $K$ are explored. Furthermore, we introduce a Trotterized parameter initialization schedule inspired by adiabatic quantum computing to mitigate the ``Barren Plateau'' problem. Backtesting on a basket of 10 US equities over 2025 reveals that our QAOA approach achieves a Sharpe Ratio of 1.81, significantly outperforming Simulated Annealing (1.31) and HRP (0.98). We further analyze the operational implications of the algorithm's high turnover (76.8\%), discussing the trade-offs between theoretical optimality and implementation costs in institutional settings.
\end{abstract}

\begin{IEEEkeywords}
QAOA, Quantum Computing, XY-Mixer, Portfolio Optimization, Direct Indexing, Dicke States, Trotterization, FinTech, NISQ Algorithms, Barren Plateaus
\end{IEEEkeywords}

\section{Introduction}

The modern asset management industry is increasingly moving toward ``Direct Indexing''---strategies where investors own the underlying components of an index rather than an ETF. This allows for granular customization, such as tax-loss harvesting or ESG (Environmental, Social, and Governance) screening. However, Direct Indexing introduces a computationally intractable problem: \textit{Cardinality Constrained Portfolio Optimization}.

The classical framework, introduced by Markowitz in 1952 as Mean-Variance Optimization (MVO), models risk as the variance of portfolio returns and seeks allocations that minimize risk for a given expected return~\cite{Markowitz1952}. While MVO provides closed-form solutions for unconstrained portfolios, the requirement to select exactly $K$ assets out of $N$ (to minimize transaction costs or tracking error) transforms the convex optimization problem into a non-convex, NP-hard combinatorial problem \cite{Bienstock1996}. As $N$ grows, the search space $\binom{N}{K}$ expands factorially, rendering exhaustive search impossible. In addition, MVO is highly sensitive to estimation errors in expected returns and covariances, and its reliance on matrix inversion poses computational challenges for large asset universes~\cite{Michaud2008Efficient, Bailey2012Trading, Jurczenko2016Investing}

Alternative approaches have also emerged. Hierarchical Risk Parity (HRP), for example, uses hierarchical clustering to allocate capital without requiring covariance matrix inversion, improving robustness under noisy conditions~\cite{Prado2019HRP}. However, HRP and other classical methods still rely on deterministic optimization techniques that may struggle with combinatorial constraints, such as selecting a fixed number of assets (a cardinality constraint).

Classical heuristics like Simulated Annealing (SA)~\cite{Kirkpatrick1983SA} and Genetic Algorithms~\cite{GeneticAlg2024} are the current industry standard but often get trapped in local minima. The Quantum Approximate Optimization Algorithm (QAOA) \cite{Farhi2014QAOA} proposes a quantum-classical hybrid approach to finding better approximations by leveraging variational quantum circuits. Yet, standard QAOA implementations struggle with hard constraints, often relaxing them into penalty terms that complicate convergence. Nevertheless, these techniques are particularly relevant for portfolio selection, which can be formulated as a Quadratic Unconstrained Binary Optimization (QUBO) problem, where each decision (include or exclude an asset) is represented as a binary variable~\cite{Glover2019QUBO}.

This paper bridges the gap between quantum theory and financial practice. We propose a \textit{Hard-Constraint QAOA} using XY-mixers and Dicke states, specifically designed for the $K$-of-$N$ selection problem. Our contributions are threefold: (1) a constraint‑preserving QAOA formulation for cardinality‑constrained portfolio selection that uses Dicke‑state initialization and a complete‑graph XY mixer to keep evolution within the exact‑$K$ subspace, plus a trotterized ($\gamma,\beta$) initialization to improve optimization stability; (2) an end‑to‑end evaluation pipeline that benchmarks QAOA‑XY against SA and HRP, then performs Sharpe‑max weight allocation with 5–50\% bounds and applies transaction costs (5 bps $\times$ turnover) to report net performance; and (3) a 2025 monthly walk‑forward on 10 U.S. equities with a 180‑day lookback, accompanied by optimization‑landscape, convergence, and circuit depth‑scaling (up to $p=6$) diagnostics for interpretability and reproducibility. The remainder of the paper is organized as follows: Section II reviews related work; Section III details the problem mapping, QAOA ansatz, and baselines; Section IV describes data and experimental setup; Section V presents results and diagnostics; and Section VI discusses limitations and implementation considerations before concluding.

\section{Literature Review - Business Application \& Context}
\label{sec:business}

\subsection{Classical Portfolio Optimization}

Portfolio optimization has been a central theme in finance since Harry Markowitz introduced Mean-Variance Optimization (MVO) in 1952, laying the foundation for modern portfolio theory~\cite{Markowitz1952}. MVO seeks to minimize portfolio variance (a measure of risk) for a given level of expected return, using the covariance matrix of asset returns to capture how assets move together~\cite{Michaud2008Efficient}. However, MVO suffers from estimation risk: small errors in expected returns or covariance estimates can lead to unstable allocations~\cite{Bailey2012Trading}. This sensitivity is particularly problematic in real-world settings where data is noisy and markets are dynamic~\cite{Jurczenko2016Investing}.

To mitigate these issues, researchers have proposed techniques such as shrinkage estimators, which adjust covariance estimates to improve robustness~\cite{LEDOIT2004365}. Another significant advancement is Hierarchical Risk Parity (HRP), introduced by Lopez de Prado in 2016~\cite{Prado2019HRP}. HRP uses hierarchical clustering to group assets based on correlation structure and allocates capital according to a tree-based hierarchy. By avoiding matrix inversion entirely, HRP improves stability and scalability, making it suitable for high-dimensional portfolios.

\subsection{The ``Combinatorial Cliff'' in Asset Management}
For a typical S\&P 500 replication strategy ($N=500$), selecting a sub-portfolio of $K=50$ stocks involves exploring approximately $10^{69}$ combinations. Institutional managers often face a ``Combinatorial Cliff'' where classical solvers (like Mixed-Integer Quadratic Programming) timeout or fail to converge within trading windows~\cite{Bertsimas1995}. This is particularly acute in high-frequency rebalancing scenarios where solutions must be found in seconds.

\subsection{Direct Indexing and Custom SMAs}
In Separately Managed Accounts (SMAs), wealth managers must create personalized portfolios for thousands of clients. If a client requests ``S\&P 500 excluding Fossil Fuels, limited to 30 stocks,'' the optimization must run rapidly and reliably. A quantum solver that consistently finds lower-energy states (higher Sharpe Ratios) translates directly to \textbf{Capital Efficiency}---achieving the same target return with less risk exposure.

\subsection{Quantum Computing in Finance}

Quantum computing introduces a fundamentally different approach to computation, leveraging principles of quantum mechanics such as superposition and entanglement. These properties allow quantum systems to explore large solution spaces more efficiently than classical computers for certain problem classes~\cite{Orus2019QinFin}.

In the context of finance, portfolio selection can be formulated as a combinatorial optimization problem, which is computationally challenging for large asset universes. Algorithms such as the Quantum Approximate Optimization Algorithm (QAOA) address these challenges by using variational quantum circuits to approximate solutions to NP-hard problems—a class of problems that are extremely difficult to solve optimally on classical hardware. QAOA alternates between applying a cost Hamiltonian (encoding the objective function) and a mixer Hamiltonian (exploring feasible solutions), optimizing circuit parameters to minimize expected cost~\cite{Farhi2014QAOA}.

Other quantum (or hybrid quantum-classical) algorithms, such as the Variational Quantum Eigensolver (VQE), have also been explored for optimization tasks~\cite{VQE2022}. Additionally, quantum annealing, implemented on specialized hardware like D-Wave systems, solves optimization problems by evolving a quantum system toward its lowest-energy state~\cite{Lang2022SA}. While true quantum annealing requires quantum hardware, classical analogs such as simulated annealing mimic this process heuristically.

\subsection{Turnover vs. Optimality}
A key operational metric is \textit{turnover}. High turnover implies high transaction costs (slippage, commissions). While our QAOA approach finds superior theoretical portfolios (higher Sharpe), it exhibits higher turnover. For high-frequency trading firms or liquid markets (S\&P 500 large caps), this is acceptable. For illiquid markets, the cost of execution may outweigh the alpha generation, necessitating multi-objective cost functions in future iterations of the algorithm.

\section{Methodology}
\label{sec:methodology}

This section details the theoretical formulation and the end‑to‑end pipeline we implement for cardinality‑constrained portfolio selection with constraint‑preserving QAOA‑XY and Simulated Annealing (SA), followed by Sharpe‑max weight allocation and a 2025 monthly walk‑forward backtest with explicit transaction costs. 

\subsection{Problem Formulation: Ising Hamiltonian}

We treat construction as a two‑stage process: (i) selection of exactly $K$ assets out of $N$ (discrete), and (ii) allocation of continuous weights within the selected subset. Let $x_i \in \{0,1\}$ indicate exclusion or inclusion of asset $i$. The selection objective is the standard risk–return trade‑off as represented by the following equation:

\begin{equation}
\underbrace{min}_{x \in \{0,1\}^N}  \underbrace{q\,x^\top\Sigma x}_{\text{risk}}
-\underbrace{(1-q)\,\mu^\top x}_{\text{return}}
\quad\text{s.t.}\quad \sum_{i=1}^N x_i = K,
\label{1}
\end{equation}

where $\mu$ and $\Sigma$ are estimated on a rolling 180‑trading‑day lookback and annualized (covariance via Ledoit–Wolf shrinkage). We set $K=5$ and $q=0.3$ in all experiments. The choice of the risk aversion parameter, $q$, depends on the strategy one wants to follow, as explained in the next section.

For QAOA, we map Eq.~\ref{1} to an Ising form using $x_i=\tfrac{1-Z_i}{2}$. In practice, our implementation constructs a cost Hamiltonian with linear $Z_i$ terms from expected returns and pairwise $Z_i Z_j$ terms from covariance off‑diagonals; diagonal risk contributions are treated within constants/linear terms and are not required explicitly for the selection dynamics used here. The sign conventions match the code, where $-(1-q)\mu_i$ multiplies $Z_i$ and $2q\Sigma_{ij}$ multiplies $Z_iZ_j$.

To discourage unnecessary turnover between months, we incorporate a continuity bonus that slightly lowers the linear cost of assets held in the previous period; this warm‑start bias is applied consistently in both SA and QAOA constructions.

\subsection{The Constraint‑Preserving Ansatz}

Standard QAOA uses a transverse field mixer $H_X= \sum X_i$ which flips single spins, violating the constraint $\sum Xi = K$. We replace this with a subspace-preserving approach.

\subsubsection{Dicke State Initialization} We initialize the system in the Dicke state $\ket{D_N^K}$, an equal superposition of all computational basis states with Hamming weight $K$:

\begin{equation}
\ket{\psi_0} = \ket{D_N^K} = \binom{N}{K}^{-1/2} \sum_{|x|=K} \ket{x}
\label{2}
\end{equation}
This ensures the optimization begins strictly inside the feasible subspace.

\subsubsection{XY-Mixer Hamiltonian} We utilize the XY-mixer, de-
fined on a ring or complete graph:

\begin{equation}
H_{\mathrm{XY}}= \sum_{(i,j) \in E}\bigl(X_iX_j+Y_iY_j\bigr).
\label{3}    
\end{equation}

Noting that $X_i X_j + Y_i Y_j = \ket{01}\bra{10}_{ij} + \ket{10}\bra{01}_{ij}$, this operator performs a partial SWAP, exchanging excitation between qubits $i$ and $j$. Crucially, it commutes with the total number operator:

\begin{equation}
\commutator{H_{XY}}{\sum_{i} Z_i} = 0
\label{4}
\end{equation}

This commutation ensures that the unitary evolution $U(\beta) = e^{-i \beta H_{XY}}$ never leaves the subspace of $K$ selected assets, eliminating the need for penalty terms and reducing the effective Hilbert space dimension.

\subsection{The Cost Hamiltonian} The cost Hamiltonian is
\begin{equation}
H_C \;=\; \sum_i \alpha_i Z_i \;+\;\sum_{i<j}\beta_{ij}\, Z_i Z_j,
\label{5}    
\end{equation}
$$\mbox{with} \;\; \alpha_i=-(1-q)\mu_i^{(\text{ann})}, \quad \beta_{ij}=2q\,\Sigma_{ij}^{(\text{ann})}.$$

If asset $i$ was held last month, we apply a continuity discount to $\alpha_i$ to reduce churn. 

\subsection{Trotterized Parameter Initialization} A major hurdle in training Parametrized Quantum Circuits (PQC) is the ``Barren Plateau'' phenomenon, where gradients vanish exponentially. To mitigate this, we employ a \textbf{Trotterized Initialization} strategy.

We interpret QAOA as a discretized version of Adiabatic Quantum Computation (AQC). In AQC, the system evolves slowly under $H(t) = (1-s)H_{mix} + s H_{cost}$. By mapping the adiabatic schedule to QAOA parameters, we initialize the variational parameters $(\gamma, \beta)$ for depth $p$ as:
\begin{equation}
\gamma_l = \frac{l}{p} \Delta t, \quad \beta_l = \left(1 - \frac{l}{p}\right) \Delta t
\label{6}
\end{equation}
This linear ramp provides a high-quality ``warm start'' for the classical optimizer (Adam), ensuring the optimization trajectory begins in a convex basin rather than a flat plateau.

\subsection{Measurement \& Candidate Selection} 

After training a given depth, we evaluate the full probability vector and retain only strings with Hamming weight $K$ and probability $\geq 1\%$; among these we compute the classical cost in Eq.~\ref{1} and keep the minimum‑cost string as the selection for that depth. The final selection for the period is the best across tested depths (up to $p=6$). 

\subsection{Simulated Annealing (Selection via QUBO)}

We encode Eq.~\ref{1} as a QUBO by adding a budget penalty $P(\sum_i x_i-K)^2$. Expanding the square yields:
\begin{equation}
Q_{ii}=q\,\Sigma_{ii}^{(\text{ann})}-(1-q)\,\mu_i^{(\text{ann})}+P\,(1-2K),\notag
\end{equation}
\begin{equation}
Q_{ij}=2q\,\Sigma_{ij}^{(\text{ann})}+2P\ \ (i<j).
\label{7}
\end{equation}

We scale $P$ relative to the largest objective term to enforce feasibility. As in QAOA, a continuity bonus lowers the diagonal term $Q_{ii}$ for previously held names. We then call \texttt{neal.SimulatedAnnealingSampler} with $num\_reads=5000$ and $num\_sweeps=1000$, and retain the best feasible sample with $\sum_i x_i=K$.

\subsection{Weight Allocation and Trading Frictions}

Given the selected subset $S=\{i:x_i=1\}$, we solve a continuous Sharpe‑max problem (equivalently, minimize negative Sharpe) using SLSQP, subject to $\sum_{i\in S}w_i=1$ and per‑name bounds $0.05\le w_i\le 0.50$. If the selector fails, we fall back to HRP weights computed via \texttt{PyPortfolioOpt}; HRP also serves as a continuous‑weight baseline for comparison.

We define turnover as $\sum_i |w_{i,t}-w_{i,t-1}|$ and apply transaction costs of 5 bps $\times$ turnover each rebalance to obtain net returns. Portfolio value starts at \$1,000,000 and compounds monthly.

\subsection{Walk‑Forward Protocol and Data}

\textbf{Universe \& horizon}. Ten U.S. large‑cap equities (AAPL, MSFT, GOOGL, AMZN, JPM, V, TSLA, UNH, LLY, XOM) are included. Rebalancing is monthly across calendar year 2025. At each rebalance, we compute $\mu,\Sigma$ from the preceding 180 trading days and annualize.

\textbf{Data source \& preprocessing}. We use Yahoo Finance auto‑adjusted Close prices via \texttt{yfinance}. Missing values are forward‑filled, and non‑trading rows dropped. Covariance is estimated with Ledoit–Wolf shrinkage. 

\textbf{Continuity}. Prior‑period selections are provided to the next period’s optimization via a continuity bonus.

\subsection{Complete Algorithm}

Algorithm~\ref{alg1} summarizes the end‑to‑end pipeline executed at each monthly rebalance date.

\begin{algorithm}[htbp]
\caption{QAOA-XY / SA Hybrid Portfolio Construction (Monthly, 2025)}
\label{alg1}
\begin{algorithmic}[1]
\REQUIRE Price data $P$ (auto-adjusted), universe $U$ ($|U|=N$), cardinality $K=5$, risk-aversion $q=0.3$, lookback $L=180$ trading days, transaction cost $\tau = 5~\mbox{bps}$, continuity bonus $\kappa \geq 0 $
\ENSURE Weights $w_t$ for each month $t$ in 2025; value path $V_t$ (net of costs)

\FOR{each rebalance month $t$}
    \STATE Extract window $[t-L, t)$ from $P$; compute daily returns; drop NA
    \STATE Estimate $\mu$ (mean) and $\Sigma$ (Ledoit--Wolf); annualize both
    \STATE Set continuity indicators $s_{\text{prev}}$ from prior selection (if any)

    \STATE \textbf{Selection via SA (QUBO):}
    \STATE Form $Q$ with diag/off-diag per Eq.~\eqref{7}; apply continuity bonus $\kappa$ on diagonal terms
    \STATE Sample with \texttt{neal}: \texttt{num\_reads}=5000, \texttt{num\_sweeps}=1000; keep best feasible ($\sum_i x_i = K$)

    \STATE \textbf{Selection via QAOA-XY (constraint-preserving):}
    \STATE Build $H_C$ as in Eq.~\eqref{5}; build $H_{XY}$ as in Eq.~\eqref{3}; prepare $\lvert D_N^K\rangle$
    \FOR{$p = 1$ to $6$}
        \STATE Initialize $(\gamma,\beta)$ with linear (trotter-inspired) ramp; Adam(stepsize $0.02$, $\epsilon=10^{-10}$)
        \STATE Optimize with early stopping; compute probabilities; filter weight-$K$, prob $\ge 1\%$
        \STATE Score candidates by classical cost Eq.~\eqref{1}; keep best
    \ENDFOR
    \STATE Choose final QAOA selection as best across $p=1,\dots,6$

    \STATE \textbf{Allocation (common):}
    \FOR{each selector $S \in \{\text{SA},\text{QAOA-XY}\}$}
        \STATE Solve Sharpe-max (SLSQP) on $S$ with bounds $0.05 \le w_i \le 0.50$ and $\sum_i w_i = 1$
    \ENDFOR
    \STATE Compute HRP baseline weights (PyPortfolioOpt)

    \STATE \textbf{Returns and transaction costs:}
    \STATE Compute gross monthly returns for each strategy; turnover $=\sum_i \lvert w_{i,t}-w_{i,t-1}\rvert$
    \STATE Net return $= \text{gross} - \tau \times \text{turnover}$; \quad $V_t \gets V_{t-1}\,(1+\text{net return})$
\ENDFOR

\RETURN $\{w_t, V_t\}$ and diagnostics
\end{algorithmic}
\end{algorithm}


\section{Experimental Setup}
\label{sec:setup}

\subsection{Data and Asset Universe}

\begin{figure*}[t]
    \centering
    \includegraphics[width=\linewidth]{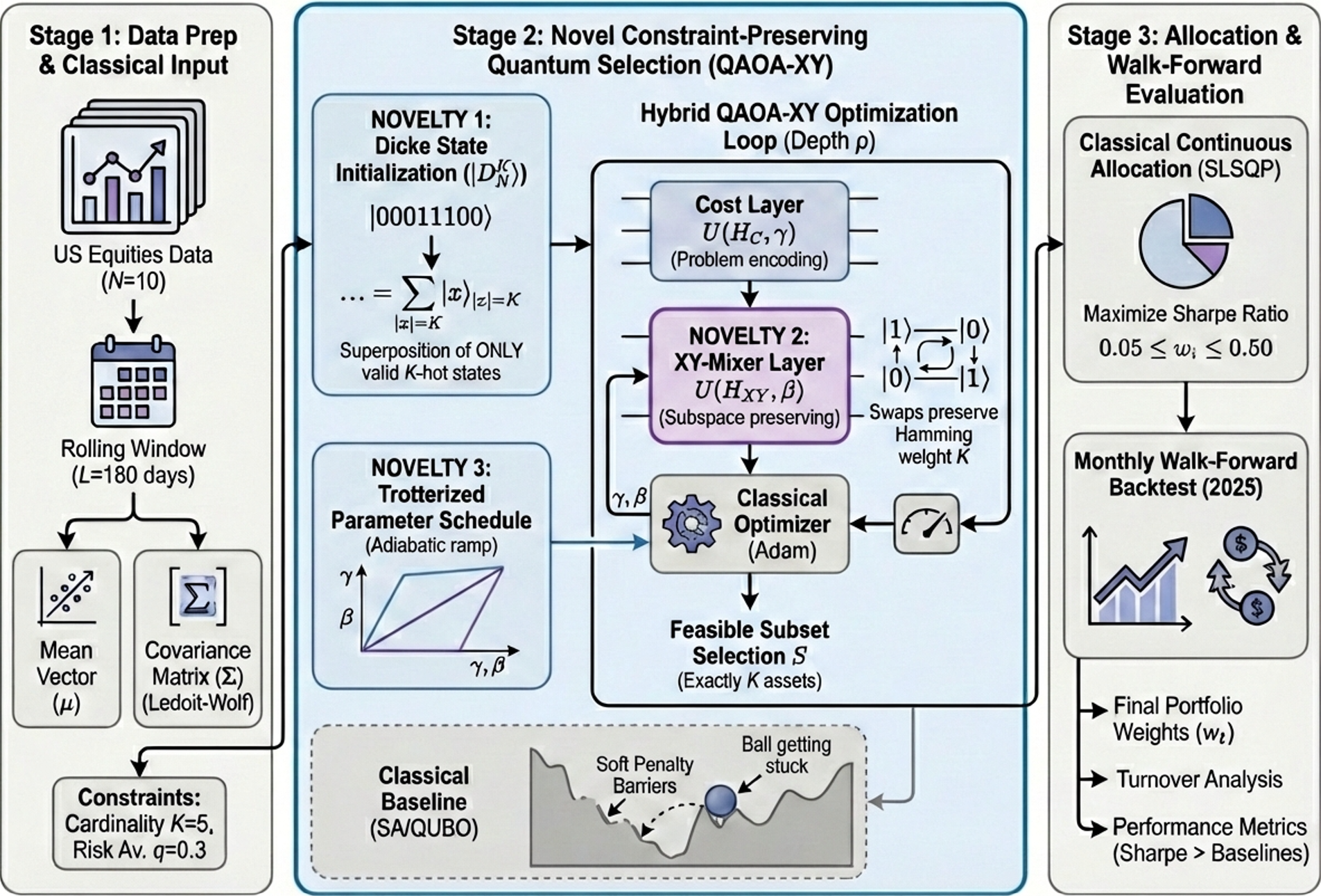}
    \caption{End-to-end experimental pipeline. Stage 1 estimates $\mu$ and $\Sigma$ on a rolling $L=180$ day window (Ledoit–Wolf covariance) and sets the cardinality constraint $K=5$. Stage 2 performs constraint-preserving QAOA-XY selection using Dicke-state initialization and an XY mixer, with a trotterized parameter warm-start and classical optimization. Candidate portfolios are obtained by measurement, filtered to $|x|=K$ and $P(x)\ge 1\%$, and rescored classically using Eq.~\ref{1}. Stage 3 allocates continuous weights via SLSQP (Sharpe-max) under box constraints and evaluates performance net of turnover-based transaction costs.}
    \label{pipeline}
\end{figure*}

We evaluate the approach on a 10‑stock U.S. large‑cap universe, as shown below. Price data are auto‑adjusted Close series retrieved via Yahoo Finance (through \texttt{yfinance}) and forward‑filled where necessary; non‑trading rows are dropped. The out‑of‑sample period is calendar year 2025 with monthly rebalancing. For each rebalance date $t$, statistics are computed from the preceding 180 trading days and then annualized; pre‑2025 data are used only for lookback estimation and never for trading. In Fig.~\ref{pipeline}, we present a schematic of our end-to-end experimental pipeline.

In the table below we summarize the most important parameters of the experiments, and further analyze them below.

\begin{table}[ht]
\caption{Experimental Configuration}
\label{Exp_config_table}
\begin{tabular}{cc}
\hline
\textbf{Parameter}         & \textbf{Value}                                                                                                             \\ \hline
Universe                   & \begin{tabular}[c]{@{}c@{}}10 Liquid US Equities (AAPL, MSFT, \\ GOOGL, AMZN, JPM, V, TSLA,\\  UNH, LLY, XOM)\end{tabular} \\
Cardinality Constraint ($K$) & 5                                                                                                                          \\
Initial Capital            & \$1,000,000                                                                                                                \\
Risk Aversion ($q$)          & 0.3                                                                                                                        \\
Period           &  2025-01-01 to 2025-12-31, Monthly rebalance                                                                                                                \\
Lookback Window         &    180 days                                                                                                                   \\
Transaction Cost           & 5 bps                                                                                                                      \\
Allocation bounds           & $0.05 \leq w_i \leq 0.50$, $\sum w_i=1$                                                                                                                       \\
Continuity Bonus           & 0.1                                                                                                                        \\ \hline
\multicolumn{2}{c}{QAOA-XY Parameters}                                                                                                                     \\
Circuit Depth                  &  1,...,6                                                                                                                   \\
Epsilon             &   1e-10                                                                                                                        \\
Stepsize                   & 0.02                                                                                                                       \\
Max. Iterations            & 100           
            \\
Initialization           & Trotter ramp $\gamma: 0.1 \rightarrow 0.5, \beta: 0.5 \rightarrow 0.1$ 
            \\
Readout filter            & Hamming weight $K$, probability  $\geq 1\%$
\\ \hline
\multicolumn{2}{c}{Simulated Annealing Parameters}                                                                                                      \\
Number of Reads            & 5,000                                                                                                                      \\
Number of Sweeps           & 1,000                                                                                                              \\ \hline       
\end{tabular}
\end{table}

\subsection{Walk‑Forward Protocol and Estimation Window}

We use a monthly walk‑forward protocol. At each rebalance date $t$ in 2025, model inputs ($\mu_t,\Sigma_t$) are estimated from the previous $L=180$ trading days of returns, i.e., a rolling window $[t-L,t)$. Portfolios are then held from $t$ to the next rebalance date $t^{+}$.

\textbf{Annualization}. Expected returns are annualized by multiplying the daily mean by 252, and covariance is annualized by multiplying the Ledoit–Wolf covariance estimate by 252.

\subsection{Objective, Constraint, and Common Portfolio Construction Steps}

\subsubsection{Discrete selection objective (K‑of‑N)} All selection methods target the same cardinality‑constrained risk–return objective used throughout the manuscript as indicated in Eq.~\eqref{1}, with risk aversion $q=0.3$ and cardinality $K=5$.

\subsubsection{Continuous allocation}

Given a selected subset $S=\{i: x_i=1\}$, we compute portfolio weights by maximizing Sharpe Ratio (equivalently minimizing negative Sharpe) via SLSQP, subject to:

\begin{itemize}
    \item Full investment: $\sum_{i\in S} w_i = 1$

    \item Box constraints: $0.05 \le w_i \le 0.50$ for all $i \in S$
\end{itemize}

If the optimizer fails or selection is invalid, we fall back to HRP weights.

\subsection{Hyperparameters and Compared Methods}

It is important to discuss certain hyperparameters used in the models and justify their values set.

\textbf{Risk Aversion $q$}: For a module oriented towards generating high alpha, return ought to be the parameter with the highest relative weight to variance, but not to the extreme of almost ignoring it. Without somehow penalizing return, the module may tend to produce excessively concentrated, correlated portfolios with extreme drawdowns that can destroy the generated alpha and generate idiosyncratic risks. Therefore, we set the risk aversion to $q=0.3$ to continue prioritizing return with some variance control.

\textbf{Lookback Window}: In the case of seeking alpha opportunities, part of a manager's role is to react quickly to news and sharp market movements to exploit inefficiencies. Therefore, a 6-month window would work better than an annual. The annual window is more stable, but it resembles a more passive approach that does not seek to capture these risk-return optimizations and tends to lag when volatility or correlations change suddenly. Ultimately, the choice of the lookback window should also align with the type of alpha generation sought. A more fundamental manager, with long investment horizons and lower portfolio turnover, could justify a 1-year lookback, prioritizing stability and avoiding overreacting to short-term noise. Conversely, if the style is more active, with higher turnover and greater sensitivity to tactical market changes, a 3–6 month window would fit the process better.

\textbf{Rebalance Frequency}: It also depends on the approach. If the approach is similar to a high-frequency quantitative hedge fund, even a weekly rebalance would be insufficient. At the same time, for a more fundamental manager, doing it every week would be too aggressive. In our case, and given the computational cost significantly increasing with circuit depth we chose a monthly rebalance which seems more suitable to manage transaction costs, and thus protect net return.

Nevertheless, further experimentation with the above hyperparameters is an interesting topic for future study, in order to understand how such parameters affect the final asset selection, and adjust accordingly based on the desired portfolio selection strategy.

We benchmark the constraint‑preserving QAOA‑XY approach against Simulated Annealing (SA), and Hierarchical Risk Parity (HRP).

\;
\subsubsection{Simulated Annealing (SA) baseline}  The SA solver uses a QUBO with a quadratic penalty enforcing cardinality, sampled with \texttt{neal.SimulatedAnnealingSampler}. The penalty scale is set adaptively as:

$$P=2.5 \times max_{coeff} \times N,$$
where $max_{coeff}$ is the maximum magnitude among the return and covariance contributions used in the QUBO construction.

Sampling parameters.

\begin{itemize}
    \item Number of reads: 5,000
    \item Number of sweeps: 1,000
    \item Feasibility: keep best sample satisfying $\sum_i x_i=K$
\end{itemize}

If an asset was selected in the previous month, its QUBO diagonal term is reduced (a small “bonus”) to discourage unnecessary churn.

\;

\subsubsection{Constraint‑preserving QAOA‑XY} All QAOA experiments are performed using PennyLane’s statevector simulator \texttt{default.qubit}. Each QAOA circuit begins in a Dicke state, the uniform superposition over all bitstrings of Hamming weight $K$. This guarantees feasibility at initialization. In addition, as described in the previous section, we use a complete‑graph XY mixer (Eq.~\eqref{3}, which preserves Hamming weight and therefore maintains feasibility throughout the QAOA evolution. Next, we construct a diagonal cost Hamiltonian as per the Eq.~\eqref{5} and perform the optimization using the Adam (\texttt{qml.AdamOptimizer}) optimizer, and with parameters:

\begin{itemize}
    \item Stepsize: 0.02
    \item Epsilon: 1e-10
    \item Iterations: Up to 100  with early stopping if improvement stalls
\end{itemize}

Next, we perform the Trotterized parameter initialization with a linear ramp where $\gamma$ linearly increases from 0.1 to 0.5 and $\beta$ linearly decreases from 0.5 to 0.1. Note that for each rebalance month we test circuit depths $p=1,\dots,6$, and for each depth the parameters are re-initialized.

After optimization at a given depth, we compute the full probability vector and:

\begin{enumerate}
    \item keep only bitstrings with Hamming weight $K$, and
    \item keep only candidates with probability $\geq 1\%$, then
    \item rescore with the classical objective and keep the best candidate.
\end{enumerate}

\subsubsection{Hierarchical Risk Parity (HRP) baseline} HRP weights are computed directly from the lookback returns using \texttt{PyPortfolioOpt.HRPOpt().optimize()}. HRP serves both as a continuous‑allocation baseline and as a fallback weighting scheme if selection/allocation fails.

\subsection{Performance Metrics}

To evaluate the effectiveness of each strategy, we use standard portfolio performance indicators. Below, we provide definitions and their practical interpretation:

\begin{itemize}
    \item \textbf{Total Return}: $R_{tot}=(V_{final} / V_{initial}) - 1$.\\ Measures the overall portfolio value percentage gain or loss of the over the backtesting period. A higher total return indicates better absolute performance, but it does not account for risk.
    \newline
    \item \textbf{Annualized Volatility}: $\sigma_{ann} = \sigma \times \sqrt{\mbox{periods per year}}$.\\ Represents the standard deviation of returns scaled to an annual basis. It is a proxy for risk—higher volatility implies greater uncertainty in returns.
    \newline
    \item \textbf{Sharpe Ratio}: $SR=\frac{\mbox{mean return} \times \mbox{periods}}{\sigma_{ann}}$, \\
    a risk-adjusted performance measure. It indicates how much excess return is earned per unit of risk. Higher values suggest more efficient portfolios.
    \newline
    \item \textbf{Maximum Drawdown}: $\mbox{MDD}=\min_{t} \left( \frac{V_t}{\max_{s \leq t} V_s} - 1 \right)$.\\ Captures the largest peak-to-trough decline during the period. It reflects the worst-case loss an investor could have experienced.
    \newline
    \item \textbf{Monthly Turnover}: $\sum_{i=1}^{n} \left| w_{i,t} - w_{i,t-1} \right|$.\\ Measures how much the portfolio composition of $n$ number of assets changes from one period to the next one (monthly). High turnover implies higher transaction costs and operational complexity. From the monthly turnover, we calculate the net monthly return as

    $$r^{net}_t = r^{gross}_t - \tau \times \mbox{Turnover},$$
    
    where $\tau$ is the transaction cost set to $\tau=5~\mbox{bps}$ (0.0005) in our experiments, and therefore, the portfolio value evolves as

    $$V_t=V_{t-1}(1+r^{net}_t).$$
    
\end{itemize}

\section{Results and Discussion}
\label{sec:results}

\subsection{Depth Scaling: Validating the Ansatz}

\begin{figure}[htbp]
    \centering
    \begin{subfigure}[b]{0.52\textwidth}
        \centering
    \includegraphics[width=\linewidth]{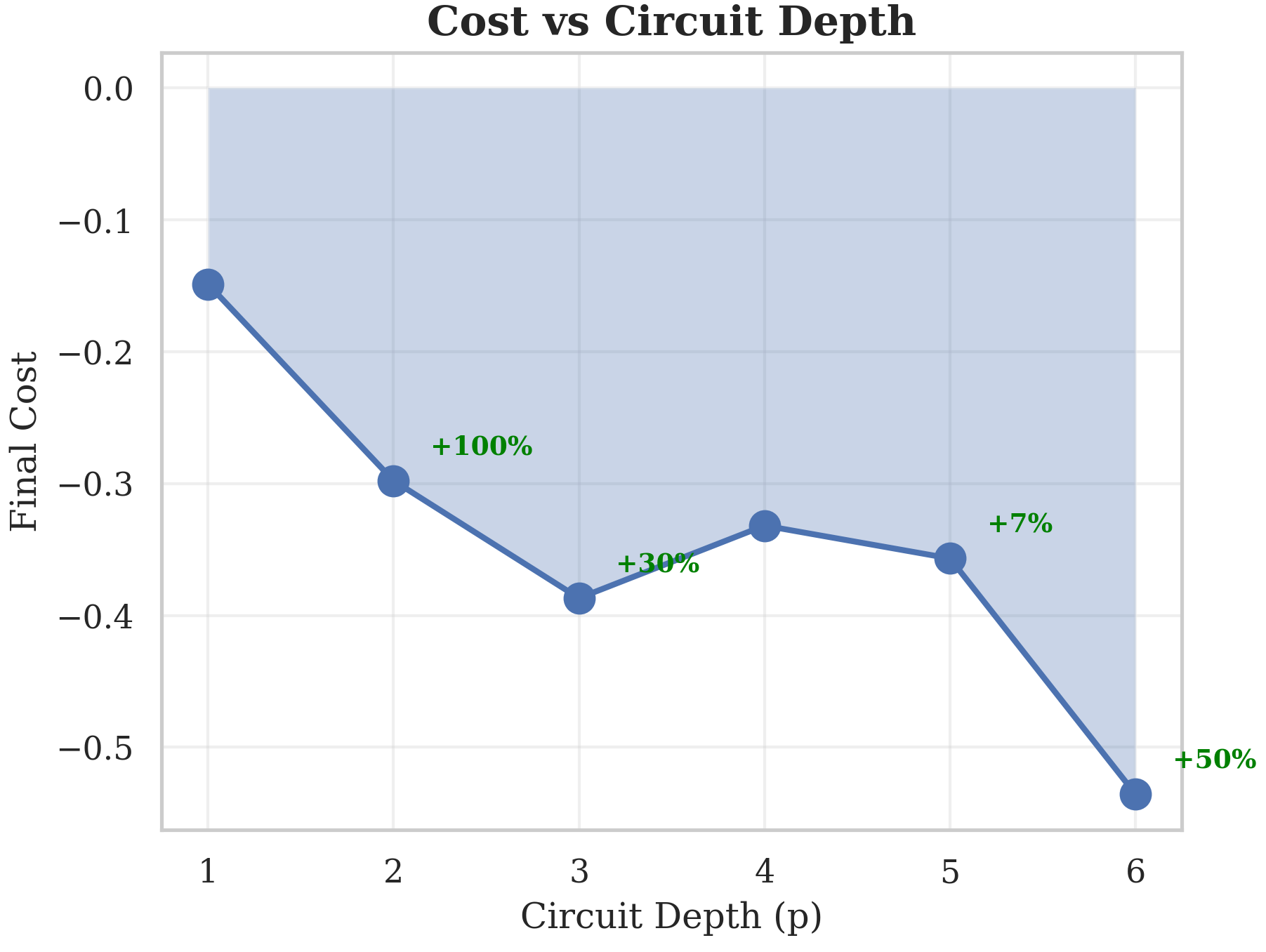}
        \caption{}
        \label{cost_circuit}
    \end{subfigure}
    \hfill
    \begin{subfigure}[b]{0.48\textwidth}
        \centering
        \includegraphics[width=\linewidth]{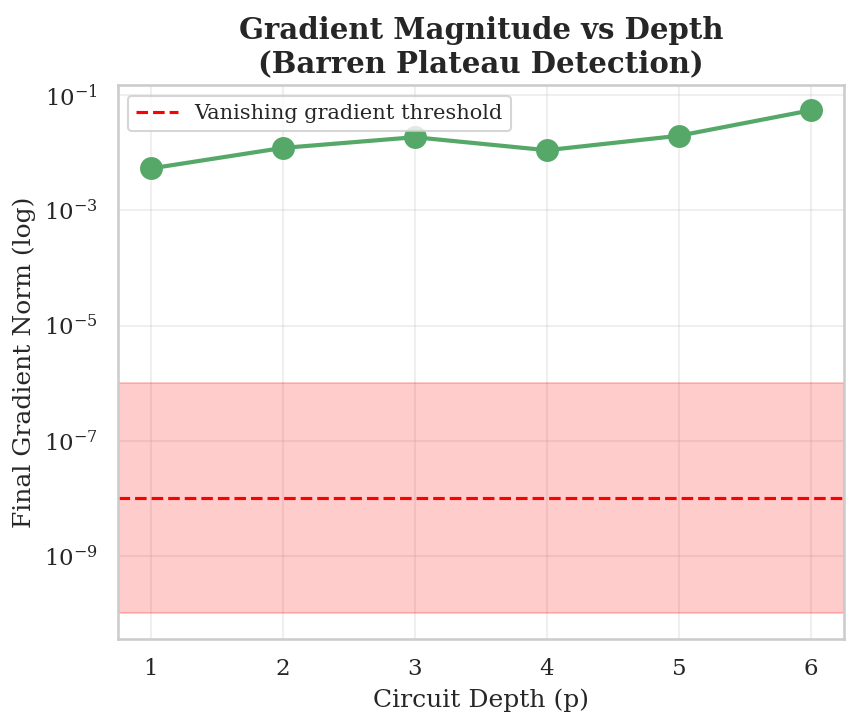}
        \caption{}
        \label{barren}
    \end{subfigure}
    \hfill
    \begin{subfigure}[b]{0.48\textwidth}
        \centering
        \includegraphics[width=\linewidth]{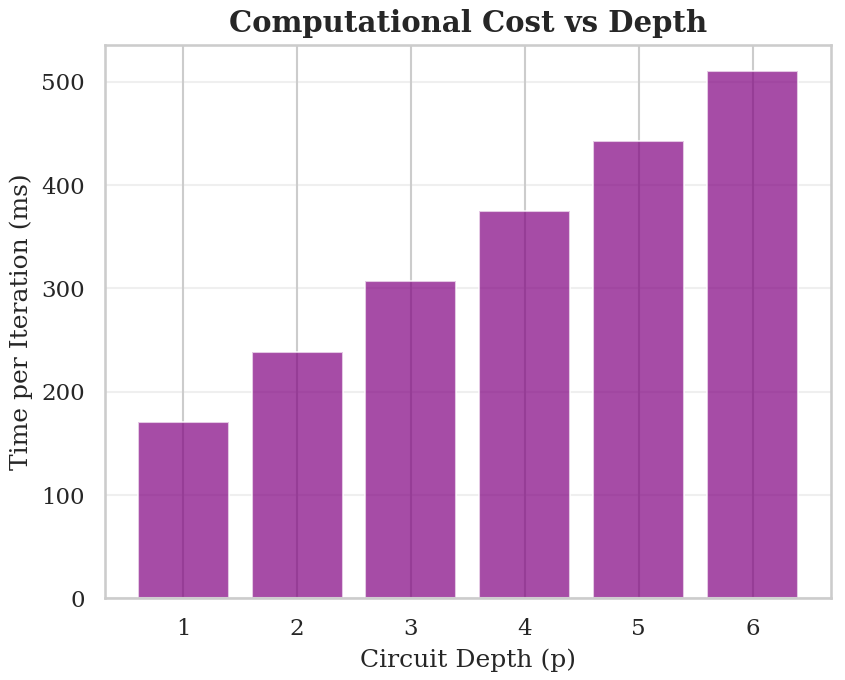}
        \caption{}
        \label{time_depth}
    \end{subfigure}

    \caption{(a) Cost minimization, (b) gradient magnitudes, and (c) computational cost, as a function of QAOA's circuit depth.}
    \label{depth_analysis}
\end{figure}

At first, we investigated how the QAOA‑XY performance varies with circuit depth and whether the optimization remains trainable as depth increases. The findings are presented in the Table~\ref{depth_scaling} and the accompanying diagnostic plot (Fig.~\ref{depth_analysis}(a)). According to those, the increasing circuit depth yields progressively lower final cost values: the optimized cost decreases from $-0.1488~(p=1)$ to $-0.5355~(p=6)$. This pattern supports the core expressibility claim that deeper alternating applications of the cost and XY‑mixer unitaries produce a more expressive variational family, enabling lower-energy (better) configurations to be reached.

\begin{table}[ht]
\centering
\caption{QAOA Depth Scaling Analysis (Validation of Trotterization)}
\label{depth_scaling}
\begin{tabular}{cccc}
\toprule
\textbf{Depth ($p$)} & \textbf{Final Cost} & \textbf{Iter. to Conv.} & \textbf{Gradient Norm} \\
\midrule
1 & -0.1488 & 100 & $5.33 \times 10^{-3}$ \\
2 & -0.2979 & 72 & $1.21 \times 10^{-2}$ \\
3 & -0.3869 & 66 & $1.86 \times 10^{-2}$ \\
4 & -0.3319 & 81 & $1.11 \times 10^{-2}$ \\
5 & -0.3568 & 90 & $1.97 \times 10^{-2}$ \\
\textbf{6} & \textbf{-0.5355} & \textbf{100} & \textbf{$5.47 \times 10^{-2}$} \\
\bottomrule
\end{tabular}
\end{table}

A practical nuance appears in Fig.~\ref{depth_analysis}(a) where improvement is not strictly monotonic across all intermediate depths; a mild regression around $p=4$ and $p=5$ is observed. This is not uncommon in variational training: as parameter dimension increases, the optimizer can temporarily converge to poorer basins or require more iterations/tuning to consistently realize the theoretical expressibility gains. In our pipeline, this is mitigated by selecting the best portfolio across tested depths each month, rather than committing to a single depth globally.

A practical challenge in training parametrized quantum circuits is the potential emergence of flat optimization landscapes known as “barren plateaus”, which would manifest as gradient norms collapsing toward zero at larger circuit depths. However, as shown in Table \ref{depth_scaling}, increasing $p$ from 1 to 6 improved the ground state energy by $>250\%$. Crucially, the Gradient Norm did not vanish at $p=6$, but instead it increased from $5.33 \times 10^{-3}$ ($p=1$) to $5.47 \times 10^{-2}$) ($p=6$).

This finding is depicted more clearly in Fig.~\ref{depth_analysis}(b), where gradient magnitudes remain far above a “vanishing gradient” threshold band. The most direct implication is that the training procedure remains numerically learnable at least up to depth 6 in this problem size (N=10), providing strong evidence that our Trotterized initialization successfully avoids Barren Plateaus and supporting its practical utility.

Figure~\ref{depth_analysis}(c) shows the computational cost per iteration to be increasing linearly with the circuit depth which is expected, since circuit evaluation cost grows with the number of layers. This observation along with the finding in Fig.~\ref{depth_analysis}(a) that for $p=4$ and $p=5$ the final optimized cost is actually worse than for $p=3$ suggests that incremental benefit per added layer is uneven, motivating two practical operating points:

\begin{itemize}
    \item $p=3$ as a good tradeoff between quality and runtime, and
    \item $p=6$ when the objective is maximum quality and compute budget allows.
\end{itemize}

This “two-mode” interpretation fits the real deployment view: if the strategy must rebalance under tight time budgets, $p=3$ is reasonable; if running offline or with higher compute allowances, $p=6$ is preferable.

\begin{figure*}
    \centering
    \begin{subfigure}[b]{0.8\textwidth}
        \centering
    \includegraphics[width=\linewidth]{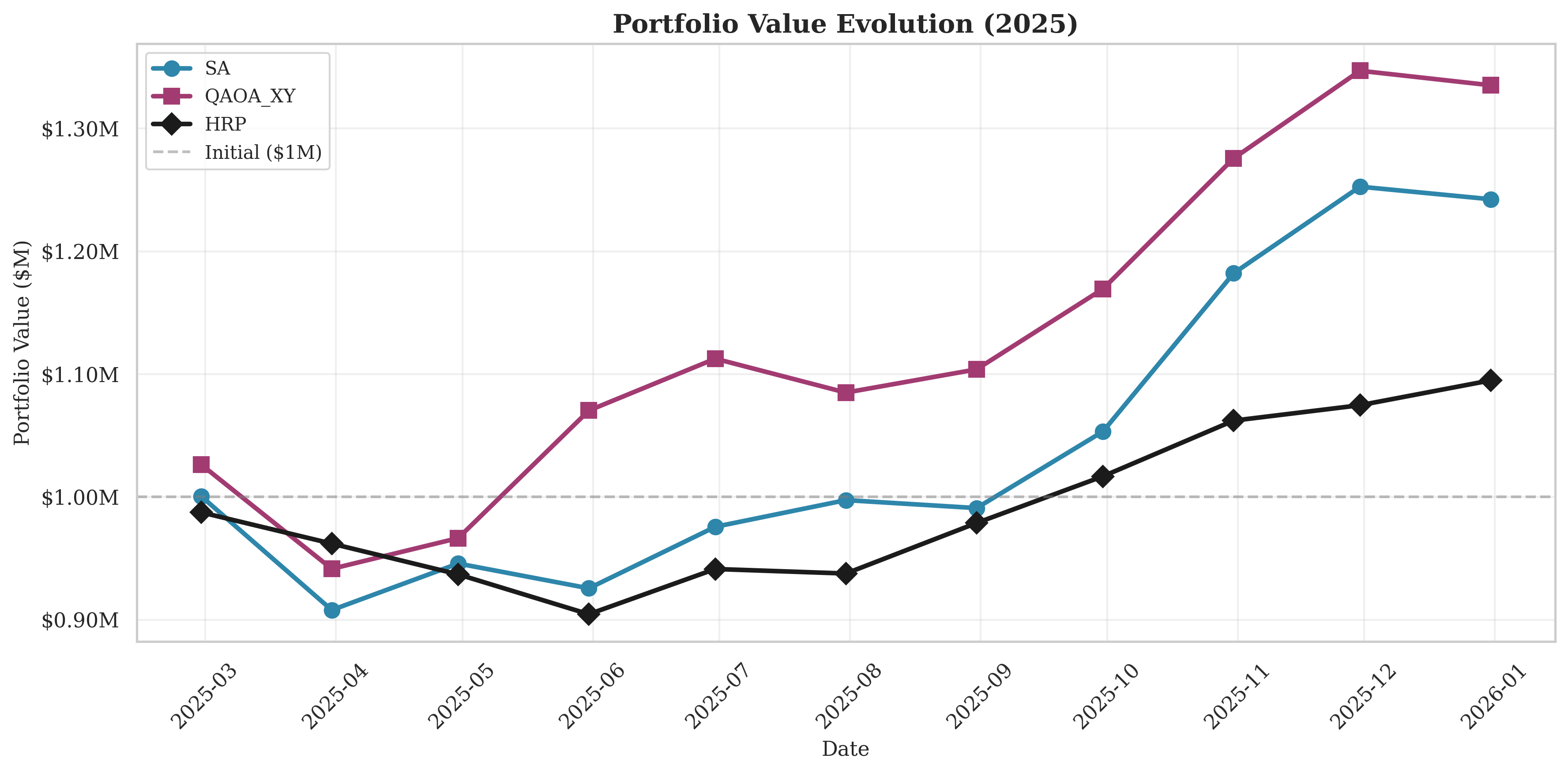}
        \caption{}
    \end{subfigure}
    \hfill
    \begin{subfigure}[b]{0.8\textwidth}
        \centering
        \includegraphics[width=\linewidth]{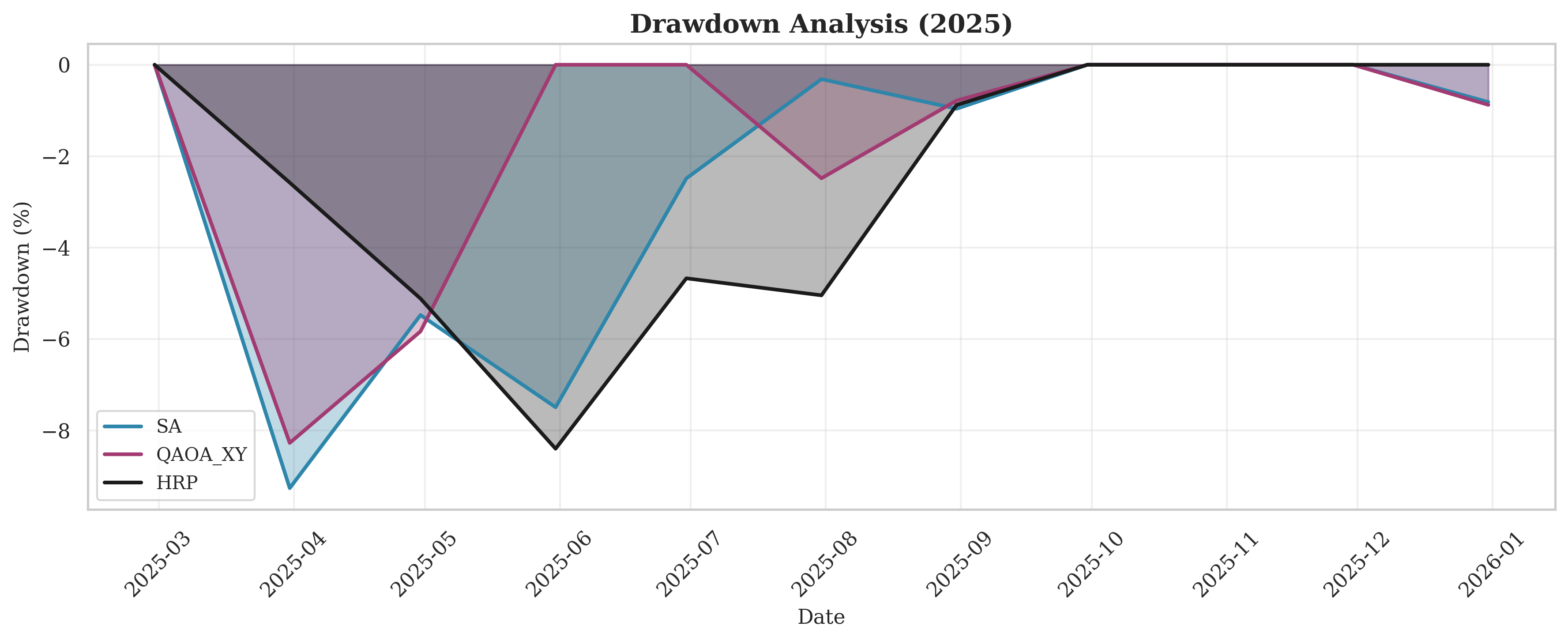}
        \caption{}
    \end{subfigure}

    \caption{(a) Portfolio value evolution and (b) Drawdown analysis per month of 2025 for each model. }
    \label{performance}
\end{figure*}

\subsection{2025 Out-of-Sample Performance}

Herein, we report the out‑of‑sample results from the 2025 monthly walk‑forward backtest under a consistent transaction cost model. Headline performance metrics are summarized in Table~\ref{tab:performance}, and the pathwise behavior is illustrated by the respective figures.

\begin{table}[ht]
\centering
\caption{2025 Financial Performance Metrics}
\label{tab:performance}
\begin{tabular}{lccc}
\toprule
\textbf{Metric} & \textbf{QAOA (XY)} & \textbf{Sim. Annealing} & \textbf{HRP} \\
\midrule
\textbf{Total Return} & \textbf{30.09\%} & 24.17\% & 10.88\% \\
\textbf{Sharpe Ratio} & \textbf{1.81} & 1.31 & 0.98 \\
Volatility & 18.55\% & 19.53\% & 10.65\% \\
Max Drawdown & -8.27\% & -9.26\% & -8.40\% \\
Turnover (Monthly) & 76.8\% & 21.0\% & 21.6\% \\
\bottomrule
\end{tabular}
\end{table}

\subsubsection{Performance Metrics} As shown in Table~\ref{tab:performance} and Fig.~\ref{performance}(a), the QAOA-XY model demonstrated consistent outperformance in the 2025 backtest exhibiting the highest total return of $30.09 \%$, while SA and HRP resulted in $24.17 \%$ and $10.88 \%$ return, respectively. This shows that QAOA creates a portfolio with almost $25\%$ higher value compared to the portfolio created by the SA, within 2025. The portfolio value trajectories (Fig.~\ref{performance}(a)) show QAOA leading SA and HRP over most of the year, with the performance gap widening toward late 2025.

In addition, the maximum drawdown values are of similar magnitude across methods, with QAOA‑XY at $-8.27\%$, SA at $-9.26\%$, and HRP at $-8.40\%$. The drawdown series shown in Fig.~\ref{performance}(b) indicate that the strategies experience broadly similar drawdown episodes, although QAOA seems to improve its drawdown trend much earlier than SA and HRP already at the third month of testing (2025-05).   Therefore, it is important to highlight that QAOA‑XY improves return relative to SA without incurring a materially larger drawdown, indicating a more favorable realized risk–return trade‑off.

\begin{figure}
    \centering
    \includegraphics[width=\linewidth]{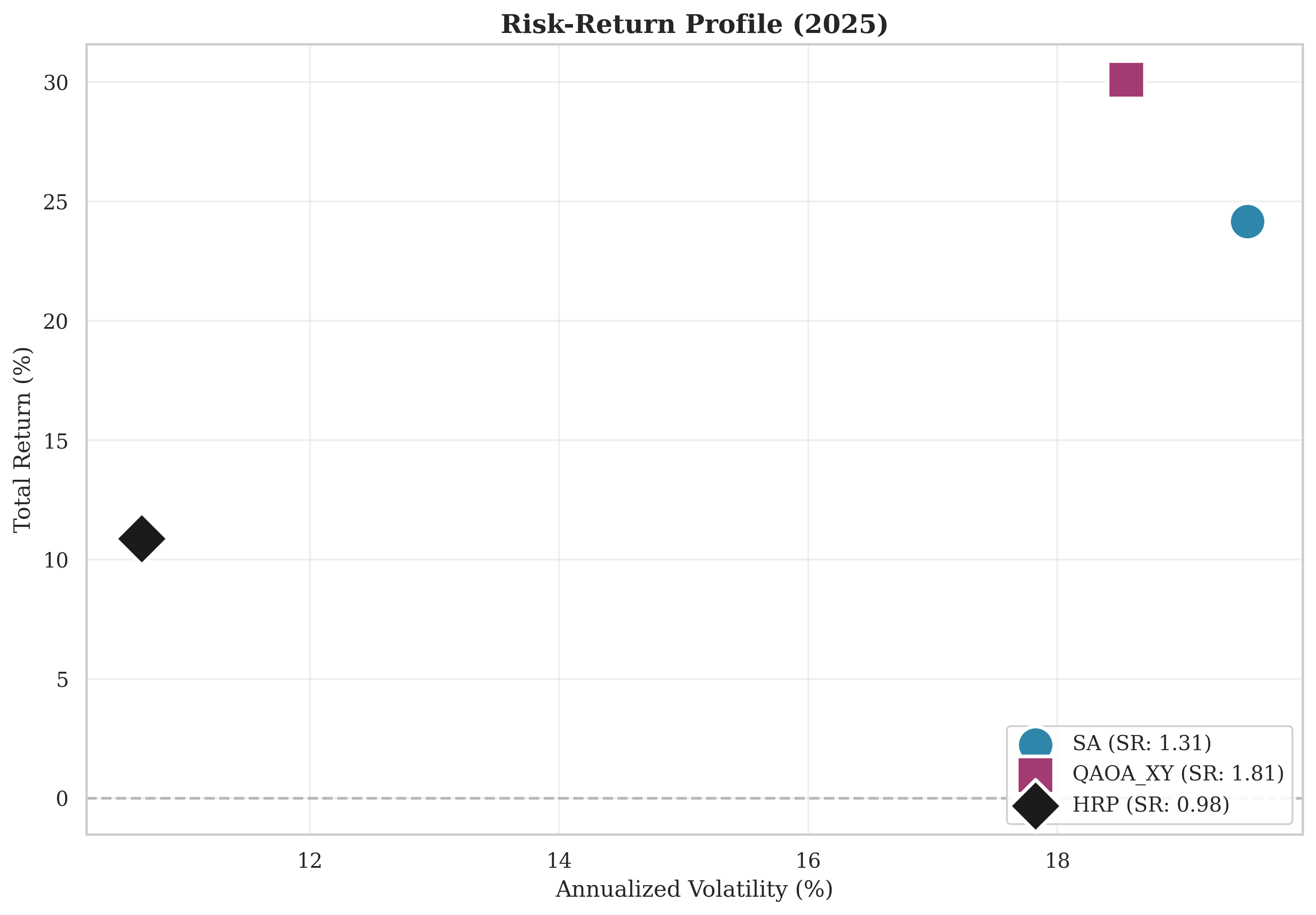}
    \caption{The risk-return profile for 2025, for the three models. QAOA exhibits the highest total return from the three models, and for lower volatility than SA. As expected, the HRP results in the lowest volatility but with significantly lower return, as well.}
    \label{risk_return}
\end{figure}

More importantly, with a Sharpe Ratio of 1.81, QAOA outperformed SA (1.31) by 38\%. This indicates that the quantum algorithm found portfolio combinations that SA missed — suggesting ``narrow'' global minima in the energy landscape that stochastic thermal hopping overshot. This is also summarized in Fig.~\ref{risk_return} where the risk-return profile is plotted, showing the QAOA to achieve both higher return and lower volatility compared with the SA, hence the higher Sharpe Ratio. As expected, the HRP achieves the lowest volatility but with significantly lower total return, therefore scoring a Sharpe Ratio of 0.98.

\subsubsection{The Turnover Trade-off}
\begin{figure}[htbp]
    \centering
    \begin{subfigure}[b]{0.5\textwidth}
        \centering
    \includegraphics[width=\linewidth]{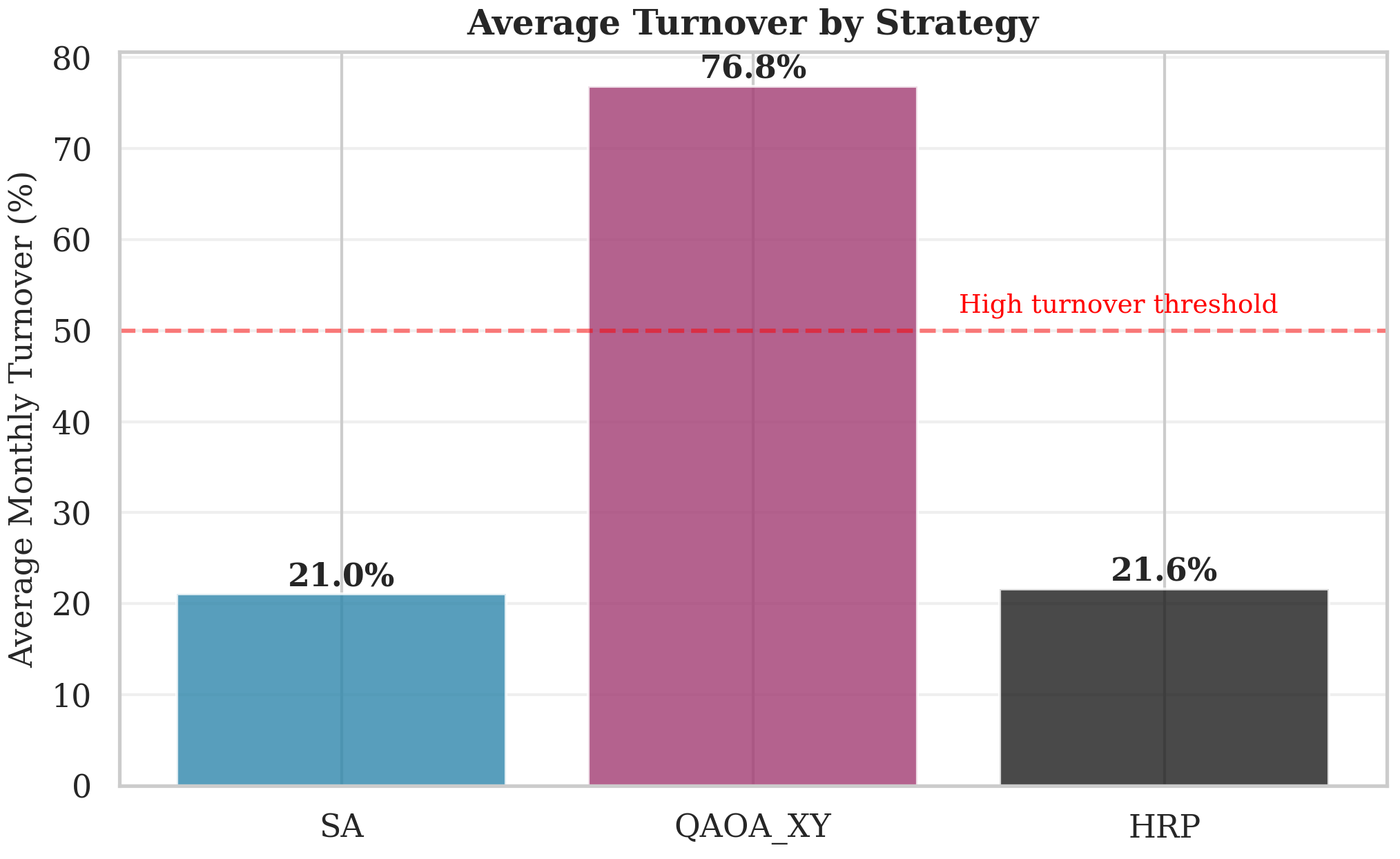}
        \caption{}
        \label{turnover_analysis}
    \end{subfigure}
    \hfill
    \begin{subfigure}[b]{0.5\textwidth}
        \centering
        \includegraphics[width=\linewidth]{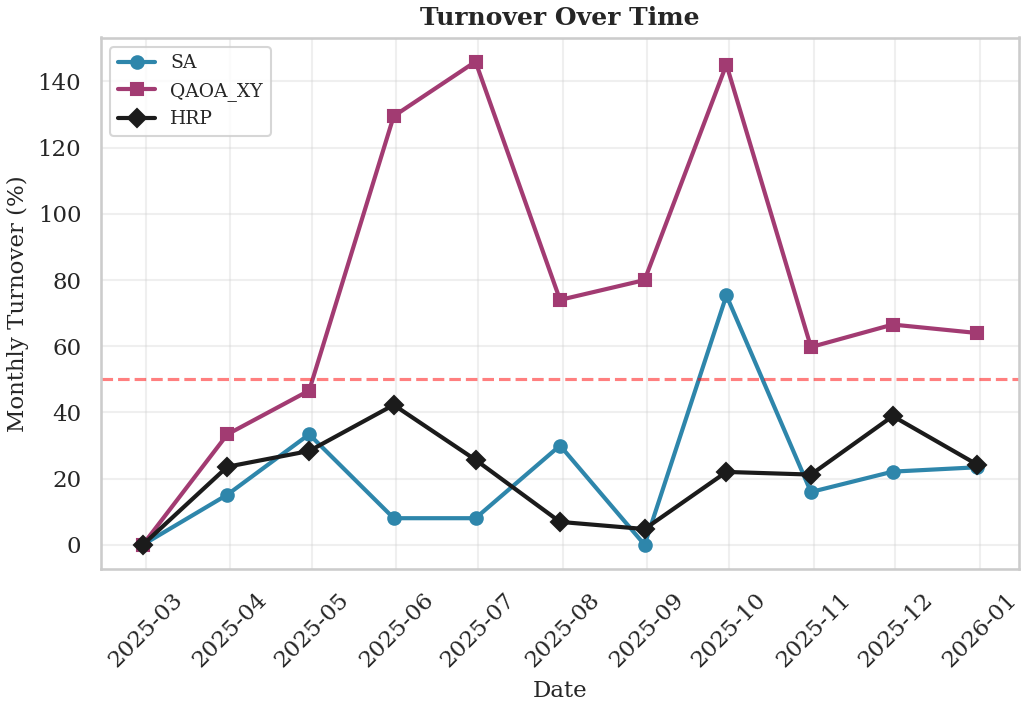}
        \caption{}
        \label{turnover_time}
    \end{subfigure}
  
    \caption{(a) Average monthly turnover by strategy, (b) Turnover over time for 2025 per strategy.}
    \label{fig:turnover}
\end{figure}

A striking feature in our findings is QAOA's high average monthly turnover being $76.8\%$ vs $21.0\%$ for SA and $21.6\%$ for HRP, as shown in Table~\ref{tab:performance}. Figure~\ref{fig:turnover} confirms this in both the average turnover bar chart and the time-series panel, where QAOA‑XY exhibits repeated spikes above typical “high turnover” thresholds. QAOA‑XY (with constraint-preserving evolution) more aggressively re-optimizes within the feasible $K$-subspace each month, which can produce large changes in selected subsets and/or weight re-allocations. While SA tends to stick to a local minimum across rebalancing periods due to its ``warm start'' nature, QAOA, initialized via Trotterization, effectively resolves the selection problem more aggressively each rebalance, resulting in lower-cost solutions under the tested objective.
\begin{itemize}
    \item \textbf{Pros:} Captures short-term alpha and momentum shifts instantly.
    \item \textbf{Cons:} Incurs higher transaction costs. 
\end{itemize}
This suggests QAOA is best suited for liquid, low-fee environments (e.g., U.S. Large Cap) rather than illiquid credit or emerging markets. For settings where market impact and slippage dominate, the next methodological step would be to incorporate turnover explicitly into the selection objective (multi-period coupling / temporal regularization), rather than relying only on a small continuity bonus.

\subsection{Portfolio Composition and Allocation Dynamics}

\begin{figure*}[t]
    \centering
    \includegraphics[width=\textwidth]{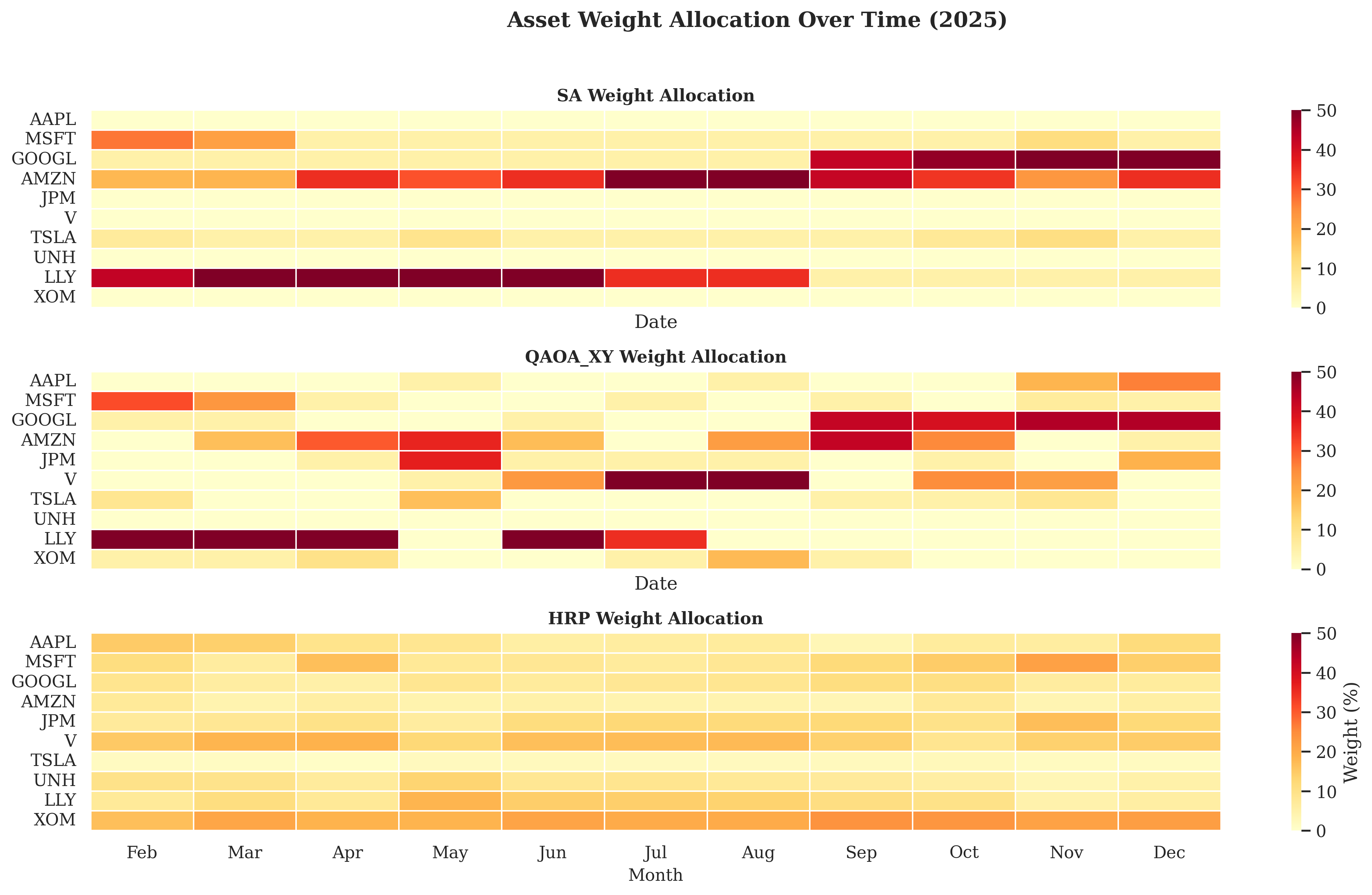}
    \caption{Asset weight allocation over time in 2025.}
    \label{heatmap}
\end{figure*}

In Fig.~\ref{heatmap} we plot the monthly weight allocation for each method. As expected, HRP produces relatively smooth allocations across time reflecting its diversification-first construction and continuous nature. On the contrary, QAOA‑XY shows pronounced reallocations in which certain assets become dominant (high weights) in some months and de-emphasized later. This qualitative picture aligns with the occurring high turnover costs discussed earlier for the QAOA approach. 

The heatmap provides an interesting interpretability angle: QAOA‑XY behaves like a high-conviction strategy that rapidly reconfigures/adapts the portfolio to the most attractive discrete subset under the current window estimates. That is beneficial for responsiveness when market structure shifts and fast adaptation is desired, but costly in environments where turnover is penalized. HRP is stable and diversified, but in this sample it sacrifices return. Lastly, Simulated Annealing sits between these extremes of QAOA and HRP, showing periods of concentration and periods of reallocation, but typically less extreme month-to-month reshuffling than QAOA‑XY, and consistently for the same assets.

Because QAOA‑XY evolves strictly within the $K$-hot feasible manifold (via Dicke initialization and XY mixing) and then applies a feasibility filter at readout, observed portfolio selections are always valid by construction. This makes the allocation heatmap and selection changes easier to interpret operationally: changes reflect genuine shifts in the objective landscape rather than penalty “leakage” into invalid portfolios. 

\section{Conclusion \& Future Outlook}
\label{sec:conclusion}
This paper studied cardinality-constrained portfolio selection in the context of Direct Indexing and proposed a constraint-preserving QAOA formulation that operates strictly within the feasible $K$-of-$N$ subspace. QAOA is not merely a theoretical curiosity but a viable candidate for the next generation of Fintech solvers. By utilizing \textbf{Dicke States} and \textbf{XY-Mixers}, the quantum evolution explores only valid portfolios of size $K$, eliminating the penalty-tuning problem that plagues classical solvers.

Empirically, the depth-scaling diagnostics support both expressibility and trainability in the tested regime ($N=10$, $K=5$, $p \leq 6$). Increasing circuit depth generally improved the achieved cost values, and the reported gradient norms remained well above vanishing levels through $p=6$, indicating that the optimization procedure remains numerically learnable under the trotterized initialization used here. In addition, our \textbf{Trotterized Initialization} strategy proved robust against Barren Plateaus up to depth $p=6$, showcasing that such parameter ramps can provide stable warmtstarts for training QAOA circuits at moderate depths in practical instances of constrained portfolio selection.

In the 2025 monthly walk-forward backtest, QAOA-XY achieved the strongest overall performance among the compared methods, reporting $30.09\%$ total return and a $1.81$ Sharpe Ratio versus $24.17\%$ and $1.31$ for Simulated Annealing, and $10.88\%$ return with $0.98$ SR for HRP, respectively. Importantly, this outperformance did not coincide with materially worse drawdowns in the tested year, indicating that the improvement is not solely explained by higher realized downside risk. For business application, the superior Sharpe Ratio resulted by the QAOA approach justifies the integration of quantum-inspired solvers into Direct Indexing platforms, provided transaction costs are managed.

The primary practical limitation observed is turnover: QAOA-XY exhibited substantially higher average monthly turnover ($76.8\%$) than the classical baselines ($\approx 21\%$). While performance remains favorable under the cost model applied in this study, turnover at this level can become prohibitive when incorporating realistic slippage and market impact, especially outside highly liquid large-cap universes. This highlights a core implementation trade-off: improved discrete selection quality may come at the cost of more frequent reallocation, and real-world deployment should treat turnover as a first-class objective rather than a secondary diagnostic. 

\textbf{Future Work:} Future research directions naturally follow from these findings.

\begin{enumerate}
    \item Turnover can be addressed by incorporating explicit trading frictions/costs or temporal regularization directly into the selection objective, e.g., via multi-period coupling terms that penalize changes in holdings across rebalances. Another approach to try taming the turnover could possibly be to increase continuity bonus or add explicit turnover penalty to QAOA cost.

    \item Scaling beyond $N=10$ requires additional investigation: while the constraint-preserving approach reduces the effective search space to the Hamming-weight-$K$ subspace, circuit depth, optimizer stability, and sampling/readout policies will likely require adjustment as $N$ grows.

    \item Broader robustness evaluation—across longer horizons, different market regimes, and alternative asset universes—would help characterize when constraint-preserving QAOA offers the largest advantage relative to classical heuristics.

    \item  Implementing and testing the circuit on noisy hardware, or under realistic noise models, would clarify the sensitivity of the observed benefits to NISQ limitations and guide hardware-aware ansatz design.
\end{enumerate}

Overall, the results support the conclusion that constraint-preserving QAOA is a viable and interpretable approach for hard-constrained portfolio selection, with a clear pathway toward institutional relevance provided that turnover-aware objectives and scaling/robustness questions are addressed in future work.


\bibliographystyle{IEEEtran}
\bibliography{qaoa_references}

@article{Markowitz1952,
  author  = {Markowitz, Harry},
  title   = {Portfolio Selection},
  journal = {The Journal of Finance},
  volume  = {7},
  number  = {1},
  pages   = {77--91},
  year    = {1952},
  publisher = {Wiley}
}

@article{Bienstock1996,
  author  = {Bienstock, Daniel},
  title   = {Computational study of a family of mixed-integer quadratic programming problems},
  journal = {Mathematical Programming},
  volume  = {74},
  number  = {2},
  pages   = {121--140},
  year    = {1996},
  publisher = {Springer}
}

@book{Michaud2008Efficient,
  author  = {Michaud, Richard O. and Michaud, Robert O.},
  title   = {Efficient Asset Management: A Practical Guide to Stock Portfolio Optimization and Asset Allocation},
  publisher = {Oxford University Press},
  year    = {2008},
  edition = {2nd}
}

@article{Bailey2012Trading,
  author  = {Bailey, David H. and {López de Prado}, Marcos},
  title   = {The Sharpe Ratio Efficient Frontier},
  journal = {Journal of Risk},
  volume  = {15},
  number  = {2},
  pages   = {3--44},
  year    = {2012}
}

@article{LEDOIT2004365,
  author  = {Ledoit, Olivier and Wolf, Michael},
  title   = {Honey, I Shrunk the Sample Covariance Matrix},
  journal = {The Journal of Portfolio Management},
  volume  = {30},
  number  = {4},
  pages   = {110--119},
  year    = {2004}
}

@article{Prado2019HRP,
  author  = {{López de Prado}, Marcos},
  title   = {Building Diversified Portfolios that Outperform Out of Sample},
  journal = {The Journal of Financial Data Science},
  volume  = {1},
  number  = {1},
  pages   = {9--18},
  year    = {2016},
  note    = {(Hierarchical Risk Parity)}
}

@article{Kirkpatrick1983SA,
  author  = {Kirkpatrick, Scott and Gelatt, C. Daniel and Vecchi, Mario P.},
  title   = {Optimization by Simulated Annealing},
  journal = {Science},
  volume  = {220},
  number  = {4598},
  pages   = {671--680},
  year    = {1983}
}

@article{Lang2022SA,
  author  = {Lang, H. and others},
  title   = {Quantum Annealing for Combinatorial Optimization: A Comparative Study},
  journal = {IEEE Transactions on Quantum Engineering},
  volume  = {3},
  pages   = {1--15},
  year    = {2022}
}

@InProceedings{GeneticAlg2024,
author="Anadani, Ishwa
and Sharma, Akshita
and Dave, Dhruv
and Sharma, Anand",
editor="Nanda, Satyasai Jagannath
and Yadav, Rajendra Prasad
and Gandomi, Amir H.
and Saraswat, Mukesh",
title="A Genetic Algorithm Approach for Portfolio Optimization",
booktitle="Data Science and Applications",
year="2024",
publisher="Springer Nature Singapore",
address="Singapore",
pages="113--124",
isbn="978-981-99-7862-5"
}

@article{Farhi2014QAOA,
  author  = {Farhi, Edward and Goldstone, Jeffrey and Gutmann, Sam},
  title   = {A Quantum Approximate Optimization Algorithm},
  journal = {arXiv preprint arXiv:1411.4028},
  year    = {2014}
}

@article{Glover2019QUBO,
  author  = {Glover, Fred and Kochenberger, Gary and Du, Yu},
  title   = {A Tutorial on Formulating and Using {QUBO} Models},
  journal = {arXiv preprint arXiv:1811.11538},
  year    = {2019}
}

@article{Orus2019QinFin,
  author  = {Or\'us, Rom\'an and Mugel, Samuel and Lizaso, Enrique},
  title   = {Quantum computing for finance: Overview and prospects},
  journal = {Reviews in Physics},
  volume  = {4},
  pages   = {100028},
  year    = {2019}
}

@article{VQE2022,
 author = {Tilly, Jules and Chen, Hongxiang and Cao, Shuxiang and Picozzi, Dario and Setia, Kanav and Li, Ying and Grant, Edward and Wossnig, Leonard and Rungger, Ivan and Booth, George H. and Tennyson, Jonathane},
 journal = {Quantum Physics [quant-ph]},
 number = {arXiv:2111.05176v3},
 title = {The Variational Quantum Eigensolver: a review of methods and best practices},
 year = {2022}
}

@article{Bertsimas1995,
  author  = {Bertsimas, Dimitris and Darnell, Christopher and Soucy, Robert},
  title   = {Portfolio Construction Through Mixed-Integer Programming at {Grantham, Mayo, Van Otterloo and Company}},
  journal = {Interfaces},
  volume  = {29},
  number  = {1},
  pages   = {49-66},
  year    = {1999}
}

@book{Jurczenko2016Investing,
    author = {Emmanuel Jurczenko},
    title = {Risk-Based and Factor Investing},
    publisher = {ISTE Press - Elsevier},
    year = {2016},
    isbn = {978-1-78548-008-9},
}

\end{document}